\documentclass{ws-procs9x6}
\usepackage{amsmath,amssymb}
\usepackage{epsfig}
\usepackage{epstopdf}

\def\eq#1{{(\ref{#1})}}

\def\dblone{\hbox{$1\hskip -1.2pt\vrule depth 0pt height 1.6ex width 0.7pt
                  \vrule depth 0pt height 0.3pt width 0.12em$}}

\newcommand{\beq}{\begin{equation}}
\newcommand{\eeq}{\end{equation}}
\newcommand{\ben}{\begin{eqnarray*}}
\newcommand{\een}{\end{eqnarray*}}

\newcommand\tr{\mathop{\mathrm{tr}}}

\begin{document}

\title{AXIAL ANOMALY, DIRAC SEA,\\
AND THE CHIRAL MAGNETIC EFFECT}

\author{Dmitri E. Kharzeev$^{1,2}$}

\address{$^1$ Department of Physics and Astronomy,\\
Stony Brook University, Stony Brook, New York 11794-3800, USA\\
$^2$ Physics Department, Brookhaven National Laboratory,\\
Upton, New York 11973-5000, USA\\
$^*$E-mail: Dmitri.Kharzeev@stonybrook.edu}

\begin{abstract}
Gribov viewed the axial anomaly as a manifestation of the collective motion of Dirac fermions with arbitrarily high momenta in the vacuum. In the presence of an external magnetic field and a chirality imbalance, 
this collective motion becomes directly observable in the form of the electric current  -- 
this is the chiral magnetic effect (CME).   I give an elementary introduction into the physics of CME, and discuss the experimental status and recent developments. 
\end{abstract}


\bodymatter

\section{Anomalies, as a manifestation of the high momentum collective motion in
  the vacuum}
  
In the article entitled ``Anomalies, as a manifestation of the high momentum collective motion in
  the vacuum,'' V.N. Gribov offered a deep insight into "one of the most beautiful and non-trivial phenomena in modern field theory" \cite{Gribov:1981ku} -- the axial \cite{Adler:1969gk,Bell:1969ts} and scale \cite{Callan:1970yg,Symanzik:1970rt,Coleman:1970je,Ellis:1970yd,Collins:1976yq} anomalies. According to Gribov, the source of anomalies can be traced back to the collective motion of particles with arbitrarily large momenta in the vacuum. This collective motion defies any UV cutoff that we may try to impose and
  "transfers the axial charge and the energy-momentum from the world with infinitely large momenta to our world of finite momenta" \cite{Gribov:1981ku}.   
  \vskip0.3cm
 Let us illustrate this statement for the case of axial anomaly by considering the Dirac sea of massless fermions. In the absence of external fields (or parity-odd interactions), the chirality is conserved and there are two disconnected Fermi surfaces of left- and right-handed fermions.  Now let us turn on an external classical field capable of changing the chirality of fermions -- e.g. the parallel electric $\vec{E}$ and magnetic $\vec{B}$ fields.  This field configuration will skew the balance between the Fermi surfaces of left- and right-handed fermions in the Dirac sea, transforming left-handed antiparticles into right-handed particles, or vice-versa, depending on the sign of the product $\vec{E}\cdot\vec{B}$. 
 \vskip0.3cm
 The mechanism of the collective flow of chirality can be described as follows\cite{Nielsen:1983rb,Witten:1984eb}:   
  the presence of magnetic field $B$ aligns the spins of the positive (negative) fermions in the direction parallel (anti-parallel) to $\vec{B}$. In the electric field $E$ the positive fermions will experience the force $e E$ and will move along $\vec{E}$; therefore their spin will have a positive projection on momentum, and we are dealing with the right fermions. Likewise, the negative fermions will be left-handed. 
After time $t$, the positive (right) fermions will increase their Fermi momentum to $p^F_R = e E t$, and the negative (left) will have their Fermi momentum decreased to $p^F_L = - p^F_R$. The one-dimensional density of states along the axis $z$ that we choose parallel to the direction of fields $\vec E$ and $\vec B$ is given by $dN_R / dz = p^F_R / 2 \pi$. In the transverse direction, the motion of fermions is quantized as they populate Landau levels in the magnetic field. The transverse density of Landau levels is $d^2 N_R/ dx dy = e B/ 2 \pi$. Therefore the density of right fermions increases per unit time as 
\beq
\frac{d^4 N_R}{dt\ dV} = \frac{e^2}{(2 \pi)^2}\ \vec{E} \cdot \vec{B}.
\eeq
The density of left fermions decreases with the same rate, $d^4 N_L / dt\ dV = -  d^4 N_R / dt\ dV$. The rate of chirality $Q_5 = N_R - N_L$ generation is thus
\beq\label{chirchange}
\frac{d^4 Q_5}{dt\ dV} = \frac{e^2}{2 \pi^2}\ \vec{E} \cdot \vec{B},
\eeq 
\vskip0.3cm
The quantity on the r.h.s. is the density of topological charge; its integral over four-dimensional space 
\beq\label{topcharge}
q[A] = {\frac{e^2}{8 \pi^2}} \int d^4 x\ F^{\mu\nu} \tilde{F}_{\mu\nu};
\eeq
reveals the  topological class to which the vector potential $A$ belongs. It has to be integer, just as the difference between the numbers of right- and left-handed fermions. 
The relation (\ref{chirchange}) thus expresses the deep connection between the axial anomaly and the topology of classical gauge fields. 
\vskip0.3cm
Having a  classical field with an infinite number of quanta is important here since the picture described above involves changing the momenta of an infinite number of particles, and "a finite number of photons is not able to change the momenta of an infinite number of particles" \cite{Gribov:1981ku}. This feature of the anomaly gives an intuitive explanation of the absence of perturbative quantum corrections to the axial anomaly that can be established formally through the renormalization group arguments \cite{Zee:1972zt,Lowenstein:1973fa,Adler:2004qt}. As will be discussed below, this property of the (electromagnetic) axial anomaly persists even when the coupling constant that determines the strength of (non-electromagnetic) interactions among the fermions becomes infinitely large. 

\vskip0.3cm
The flow of chirality, as the derivation above reveals, is accompanied by the collective motion of particles at all momenta, including the momenta around the UV cutoff scale $\Lambda_{UV}$ that we may attempt to introduce. Therefore our world of particles with finite momenta $p < \Lambda_{UV}$ cannot be isolated from the world of particles with arbitrarily high momenta, and this according to Gribov is the essence of quantum anomalies.

\section{The chiral magnetic effect and Landau levels of Dirac fermions}

Consider now the situation in which  there exists an external magnetic field, and an imbalance between the Fermi momenta of left- and right-handed fermions. In the absence of an external electric field, this imbalance cannot be caused by electromagnetic interactions, but we can imagine that the imbalance may originate from other sources -- e.g. from strong interactions of the fermions (quarks) with a non-Abelian gauge field configuration with non-trivial topological contents. 
\vskip0.3cm
The presence of magnetic field aligns the spins of positive and negative fermions in opposite directions -- along or against the direction of $\vec{B}$, respectively. Therefore being, say, right-handed means for the positive fermion 
to move along the direction of magnetic field, and for a negative fermion - to move against the direction of $\vec{B}$. Therefore, if the densities and Fermi-momenta of left- and right-handed fermions are unequal in the presence of an external magnetic field, there should be an electric current and a separation of electric charge -- this is the Chiral Magnetic Effect (CME) \cite{Kharzeev:2004ey,Kharzeev:2007tn,Kharzeev:2007jp,Fukushima:2008xe,Kharzeev:2009fn}. 
 \vskip0.3cm
Let us introduce, in addition to magnetic field $\vec{B}$, an auxiliary electric field $\vec{E}$ and consider the energy balance of chirality generation.  
Changing chirality by one unit means transferring a massless fermion from the Fermi surface of left-handed particles to the one of the right-handed particles; this change costs an amount of energy equal to the difference of the corresponding Fermi-momenta $\mu_R - \mu_L= 2 \mu_5$. If we multiply this energy by the rate of chirality change \eq{chirchange}, we get the energy spent per unit time:
\beq 
P = (\mu_R - \mu_L)\  \frac{e^2}{(2 \pi)^2}\  \int d^3 x\ \vec{E}\cdot \vec{B}.
\eeq
As we argued above this energy powers the electric current, the power of which is given by 
\beq
P = \int d^3 x\ \vec{J}\cdot \vec{E}. 
\eeq
We can take $\vec{E}$ in the direction of $\vec{B}$ in this
expression, and then get rid of the auxiliary electric field by taking the limit $\vec{E} \rightarrow 0$. 
This allows us to find the following expression for the density of CME current\cite{Fukushima:2008xe}:
\begin{equation}\label{cme}
  \vec{J}  = 
 \frac{ e^2 \mu_5}{2\pi^2} \, \vec{B}.
\end{equation}
Note that this relation manifestly violates parity since magnetic field on the r.h.s. is a pseudo-vector whereas the electric current on the l.h.s. is a vector.  Because of this, a static magnetic field with no curl cannot induce electric current in Maxwell electrodynamics (that is parity-even). In our case, the violation of parity is  induced by the imbalance between the left- and right-handed fermions. Closely related phenomena have been discussed earlier in the physics of
primordial electroweak plasma \cite{Giovannini:1997gp} and quantum
wires \cite{acf}. 
\vskip0.3cm
  
  A more rigorous derivation\cite{Fukushima:2008xe} of \eq{cme} invokes the explicit sum over the contributions of all Landau levels of charged fermions. This sum is in general divergent, and one has to introduce a UV cutoff on the energy of Landau levels -- this is the manifestation of the collective flow from the world of finite momenta to the world of infinite momenta discussed by Gribov. However, all excited Landau levels are degenerate in spin, and the opposite spin orientations give the contributions to the CME electric current that are opposite in sign and thus cancel each other. The lowest Landau level (LLL) of massless fermions is an exception since it is chiral, i.e. not degenerate in spin. Because of this, only the LLL contribution survives in the final expression \eq{cme} that does not contain any UV divergence.
  \vskip0.3cm

\section{The chiral magnetic effect and Maxwell-Chern-Simons electrodynamics}

Let us now consider the CME in the effective theory of electromagnetism obtained by integrating the quarks out of the action\cite{Kharzeev:2009fn}. Let us start from the QCD coupled to electromagnetism; the resulting theory possesses $SU(3) \times U(1)$ gauge symmetry:
$$
{\cal L}_{\rm QCD+QED} =  -{1 \over 4} G^{\mu\nu}_{\alpha}G_{\alpha \mu\nu}  + \sum_f \bar{\psi}_f \left[ i \gamma^{\mu} 
(\partial_{\mu} - i g A_{\alpha \mu} t_{\alpha} -  i q_f A_{\mu}) -  m_f \right] 
\psi_f  - 
$$
\beq\label{qcd+qed}
- {\theta \over 32 \pi^2}  g^2 G^{\mu\nu}_{\alpha} \tilde{G}_{\alpha \mu\nu} - \frac{1}{4}F^{\mu\nu}F_{\mu\nu},
\eeq 
where $A_{\mu}$ and $F_{\mu\nu}$ are the electromagnetic vector potential and the corresponding field strength tensor, and $q_f$ are the electric charges of the quarks.   
\vskip0.3cm
Let us discuss the electromagnetic sector of the theory  \eq{qcd+qed}. Electromagnetic fields will couple to the electromagnetic currents $J_\mu = \sum_f  q_f \bar{\psi}_f \gamma_\mu \psi_f$.  
In addition, the $\theta$-term in \eq{qcd+qed} will induce through the quark loop the coupling of $F \tilde{F}$ to the QCD topological charge. Let us introduce an effective pseudo-scalar field $\theta = \theta(\vec x, t)$ (playing the r\^{o}le of the axion\cite{Wilczek:1977pj,Weinberg:1977ma,Peccei:1977hh} field, but without a kinetic term) and write down the resulting effective Lagrangian as
\beq\label{MCS}
{\cal L}_{\rm MCS} = - \frac{1}{4}F^{\mu\nu}F_{\mu\nu} - A_\mu J^\mu - \frac{c}{4}\ \theta \tilde{F}^{\mu\nu} F_{\mu\nu},
\eeq
where 
\beq\label{coef}
c = \sum_f q_f^2 e^2 / (2\pi^2). 
\eeq

This is the Lagrangian of Maxwell-Chern-Simons, or axion, electrodynamics that has been introduced previously in Refs \cite{Wilczek:1987mv,Carroll:1989vb,Sikivie:1984yz}. 
\vskip0.3cm
As we discussed above, the quantity $\tilde{F}^{\mu\nu} F_{\mu\nu}$ is the density of topological charge. Therefore the integral of this quantity over a four-dimensional volume should be an (integer) topological invariant sensitive only to the long distance, global properties of the gauge field. Such properties are determined by the asymptotic behavior of the field at the surface of the four-dimensional sphere, and thus the topological invariant has to be determined by the surface integral; Gauss theorem thus dictates that $\tilde{F^{\mu\nu}} F_{\mu\nu}$ has to be a full divergence:
\beq\label{an_ab}
\tilde{F}^{\mu\nu} F_{\mu\nu} = \partial_\mu J_{CS}^\mu;
\eeq
the quantity $J_{CS}^\mu$ is the Chern-Simons current 
\beq\label{topdiv1}
J_{CS}^{\mu} = \epsilon^{\mu\nu\rho\sigma} A_{\nu} F_{\rho\sigma}, 
\eeq
that is a three-dimensional Chern-Simons form\cite{Chern:1974ft} promoted in four dimensions to a current by adding an extra index to the antisymmetric tensor. 
The Abelian three-dimensional Chern-Simons form 
\beq\label{csform}
CS[A] = \int d^3 x\ \epsilon^{\nu\rho\sigma} A_{\nu} F_{\rho\sigma}
\eeq
 is so-called magnetic helicity $\int d^3 x\ \vec{A} \cdot \vec{B}$ measuring the linkage of the lines of magnetic flux. 
\vskip0.3cm

If $\theta$ is a constant, then the entire last term in \eq{MCS} represents a full divergence -- therefore it 
does not affect the equations of motion and thus does not have any effect on the electrodynamics of charges.
The situation is different if the effective field $\theta = \theta(\vec x, t)$ varies in space-time.      
Indeed, in this case we have
\beq
\theta \tilde{F^{\mu\nu}} F_{\mu\nu} = \theta \partial_\mu J_{CS}^\mu = \partial_\mu\left[\theta J_{CS}^{\mu}\right] - \partial_\mu \theta  J_{CS}^{\mu}.
\eeq
The first term on r.h.s. is again a full derivative and can be omitted; introducing notation
\beq
P_\mu = \partial_\mu \theta = ( M, \vec P )
\eeq
we can re-write the Lagrangian \eq{MCS} in the following form:
\beq\label{CS}
 {\cal L}_{\rm MCS} = - \frac{1}{4}F^{\mu\nu}F_{\mu\nu} - A_\mu J^\mu + \frac{c}{4} \ P_\mu J^\mu_{CS}.
\eeq
Since $\theta$ is a pseudo-scalar field, $P_\mu$ is a pseudo-vector; as is clear from   \eq{CS}, 
it plays a r\^{o}le of the potential coupling to the Chern-Simons current \eq{topdiv1}. However, unlike the vector potential $A_\mu$, $P_\mu$ is not a dynamical variable and is a pseudo-vector that is fixed by the dynamics of chiral charge -- in our case, determined by the fluctuations of topological charge in QCD.
\vskip0.3cm   
Let us write down the Euler-Lagrange equations of motion that follow from the Lagrangian \eq{CS},\eq{topdiv1}  
(Maxwell-Chern-Simons equations):
\beq
\partial_\mu F^{\mu\nu} = J^\nu - P_\mu \tilde{F}^{\mu\nu}.
\eeq
The first pair of Maxwell equations (which is a consequence of the fact that the fields are expressed through the vector potential) is not modified:
\beq
\partial_\mu \tilde{F}^{\mu\nu} = J^\nu.
\eeq   
It is convenient to write down these equations also in terms of the electric $\vec E$ and magnetic $\vec B$ fields:
\beq\label{MCS1}
\vec{\nabla}\times \vec{B} - \frac{\partial \vec{E}}{\partial t} = \vec J + c \left(M \vec{B} - \vec{P} \times \vec{E}\right), 
\eeq
\beq\label{MCS2}
\vec{\nabla}\cdot \vec{E} = \rho + c \vec{P} \cdot \vec{B},
\eeq
\beq\label{MCS3}
\vec{\nabla}\times \vec{E} +  \frac{\partial \vec{B}}{\partial t} = 0,
\eeq
\beq\label{MCS4}
\vec{\nabla}\cdot \vec{B} = 0,
\eeq
where $(\rho, \vec J)$ are the electric charge and current densities.
One can see that the presence of Chern-Simons term leads to essential modifications of the Maxwell theory and induces, as we will see, the chiral magnetic effect.
\vskip0.3cm
Let us however start with a different phenomenon --  the Witten effect 
\cite{Witten:1979ey}: magnetic monopoles at finite $\theta$ angle acquire electric charge and become "dyons". 
Consider, following Wilczek \cite{Wilczek:1987mv}, a magnetic monopole in the presence of a finite $\theta$ angle. In the core of the monopole $\theta=0$, and away from the monopole $\theta$ acquires a finite non-zero value -- therefore within a finite domain wall we have a non-zero $\vec P = \vec{\nabla} \theta$ pointing radially outwards from the monopole. According to \eq{MCS2}, the domain wall thus acquires a non-zero charge density $c  \vec{\nabla} \theta \cdot \vec{B}$. An integral along $\vec P$ (across the domain wall) yields $\int dl\ \partial \theta / \partial l = \theta$, and the integral over all directions of $\vec P$ yields the total magnetic flux $\Phi$. By Gauss theorem, the flux is equal to the magnetic charge of the monopole $g$, and the total electric charge of the configuration is equal to 
\beq
q = c\ \theta\ g = \frac{e^2}{2\pi^2}\ \theta\ g = \frac{e}{2\pi^2}\ \theta\ (e g) = e\ \frac{\theta}{2 \pi},
\eeq
where we have used an explicit expression \eq{coef} for the coupling constant $c$, as well as the Dirac condition $g e = 4 \pi \times {\rm integer}$.
\vskip0.3cm
Consider now a configuration where an external magnetic field $\vec B$ pierces a domain with $\theta \neq 0$ inside;  outside $\theta=0$. Let us assume first that the field $\theta$ is static, $\dot\theta = 0$. Assuming that the field $\vec B$ is perpendicular to the domain wall, 
we find from \eq{MCS2} that the upper domain wall acquires the charge density per unit area $S$ of  \cite{Kharzeev:2007tn}
\beq
\left(\frac{Q}{S}\right)_{up}  = +\ c\ \theta B
\eeq
while the lower domain wall acquires the same in magnitude but opposite in sign charge density
\beq
\left(\frac{Q}{S}\right)_{down}  = -\ c\ \theta B
\eeq 
Assuming that the domain walls are thin compared to the distance $L$ between them, we find that 
the system possesses an electric dipole moment
\beq\label{eldip}
d_e = c\ \theta\ (B \cdot S)\ L = \sum_f q_f^2 \ \left(e\ \frac{\theta}{\pi}\right)\ \left(\frac{eB\cdot S}{2\pi}\right)\ L;
\eeq
for brevity of notations we put $\sum_f q_f^2 = 1$; it is easy to restore this factor in front of $e^2$ when needed.
Static electric dipole moment is a signature of ${\cal P}$, ${\cal T}$ and ${\cal CP}$ violation (we assume that $\cal{CPT}$ invariance holds). The spatial separation of charge will induce the corresponding electric field $\vec E = c\ \theta\ \vec B$. The mixing of pseudo-vector magnetic field $\vec B$ and the vector electric field $\vec E$ signals violation of ${\cal P}$, ${\cal T}$ and ${\cal CP}$ invariances.
\vskip0.3cm
The formula \eq{eldip} allows a simple interpretation: since $eB/2\pi$ is the transverse density of Landau levels of charged fermions in magnetic field $B$, the floor of the quantity $eB\cdot S/2\pi$ (i.e. the largest integer that is smaller than $eB\cdot S/2\pi$) is an integer number of fermions localized on the domain wall. Each fermion species contributes independently to this number as reflected by the factor $N_f$. Again we see that the electric dipole moment \eq{eldip} arises from the electric 
charge $q \sim e \theta/\pi$ that is induced on the domain walls due to the gradient of the pseudo-scalar field $\theta$.  
\vskip0.3cm
If the domain is due to the fluctuation 
of topological charge in QCD vacuum, its size is on the order of QCD scale, $L \sim \Lambda_{\rm QCD}^{-1}$, $S \sim \Lambda_{\rm QCD}^{-2}$. This means that to observe an electric dipole moment
in experiment we need an extremely strong magnetic field $eB \sim   \Lambda_{\rm QCD}^{2}$. Fortunately, such fields exist during the early moments of a relativistic heavy ion collision \cite{Kharzeev:2007jp,Skokov:2009qp}. Here we have assumed that the domain is static; this approximation requires the characteristic time of topological charge fluctuation $\tau \sim 1/\dot{\theta}$ be large on the time scale at which the magnetic field $B$ varies. This assumption is only marginally satisfied in heavy ion collisions, and so we now need to consider also the case of $\dot\theta \neq 0$. Note however that if the medium produced in heavy ion collisions conducts electricity, then the decaying with time 
magnetic field will induce the circular electric current which in accord with Lenz's law will in turn produce a magnetic field \cite{Tuchin:2010vs}. The estimates \cite{Tuchin:2010vs} indicate that this mechanism can extend the lifetime of magnetic field in a very significant way. 
 \vskip0.3cm 
 Consider now the domain where $| \vec{P} | \ll \dot{\theta}$, i.e. the spatial dependence of $\theta(t, \vec x)$ is much slower than the dependence on time \cite{Kharzeev:2007jp}. Again, we will expose the domain to an external magnetic field $\vec B$ with $\vec{\nabla}\times \vec{B} = 0$, and assume that no external electric field is present.  In this case we immediately get from \eq{MCS1} that there is an induced CME current \cite{Fukushima:2008xe}
 \beq\label{chimag}
 \vec{J} = - c\ M\ \vec{B} = - \frac{e^2}{2 \pi^2}\ \dot\theta \vec{B}.
 \eeq

\section{Chiral magnetic and chiral vortical effects\\ at strong coupling; relativistic hydrodynamics}

Many of our arguments were based on the weak coupling picture, e.g. on the existence of Landau levels obtained by solving Dirac equation in the external magnetic field. One may worry that once the strong interactions among 
the quarks are turned on, this simple picture will break down. Nevertheless this does not happen -- the essentially topological nature of the phenomenon protects it from being modified by quantum corrections, even at strong coupling.  In particular, in holographic
models (at infinite 't Hooft coupling) the magnitude of the chiral
magnetic effect \cite{Yee:2009vw,Rebhan:2009vc,Gorsky:2010xu} appears the
same as at weak coupling \cite{Rubakov:2010qi,Gynther:2010ed,Yee:2009vw}.  
\vskip0.3cm
The
CME has been studied in lattice QCD coupled to electromagnetism, both in the quenched
\cite{Buividovich:2009wi,Buividovich:2009zzb,Buividovich:2010tn} and
dynamical (domain wall) fermion \cite{Abramczyk:2009gb} formulations; these simulations fully take account of strong interactions among the (anti)quarks. This suggests that the CME exists even when the coupling among the quarks is strong.
\vskip0.3cm
Quark-gluon plasma at strong coupling has been argued to behave as a nearly perfect fluid (for review, see \cite{Schafer:2009dj}), and an effective low-energy theory of strongly interacting fluids is well known -- it is hydrodynamics. 
This invites a very interesting question about the role of axial anomaly in relativistic hydrodynamics that was addressed recently \cite{Son:2009tf}. In a fluid, the role of magnetic field at finite baryon chemical potential $\mu_B$ can be played by vorticity $\vec{\omega}$ of the local fluid velocity $\vec{v}$:
\beq
\vec{\omega} = \vec{\nabla} \times \vec{v}.
\eeq
This is quite natural since the rotating charged fluid generates an effective magnetic field $\mu_B \vec{\omega}$.
As a result, the electric current can be induced by the rotation of the fluid with finite baryon and axial charge density even in the absence of an external magnetic field \cite{Kharzeev:2007tn,Son:2009tf} -- so-called "chiral vortical effect" (CVE). 
The topological origin of this phenomenon is manifest since the analog of Chern-Simons 3-form \eq{csform} in this case is the "kinetic helicity"  $\int d^3x\  \vec{v} \cdot \vec{\omega}$. For a discussion of other vorticity-induced effects in heavy ion collisions, see e.g.  \cite{Liang:2004ph,Betz:2007kg,Becattini:2007sr}.
\vskip0.3cm
A related effect---the emergence of a chiral current in a medium with
finite baryon density, in an external magnetic field or in the
presence of a vorticity the fluid---has been introduced in Refs~\cite{Son:2004tq,Metlitski:2005pr,Son:2009tf}.
The close connection between CME and
the latter effect can be established for example by the method of dimensional
reduction appropriate in the case of a strong magnetic field
\cite{Basar:2010zd}: the simple relations $J_V^0 = J_A^1, \ J_A^0 =
J_V^1$ between the vector $J_V$ and axial $J_A$ currents in the
dimensionally reduced $(1+1)$ theory imply that the density of
baryon charge must induce the axial current, and the density of
axial charge must induce the charge current (CME).

\section{Experimental status and a new test}

Recently, STAR \cite{:2009uh, :2009txa} and PHENIX \cite{phenix,Ajitanand:2010rc}
Collaborations at Relativistic Heavy Ion Collider reported
experimental observation of charge asymmetry fluctuations.  While the
interpretation of the observed effect is still under intense
discussion, the fluctuations in charge asymmetry have been predicted
to occur in heavy ion collisions due to the Chiral Magnetic Effect
(CME) in QCD coupled to electromagnetism
\cite{Kharzeev:2004ey,Kharzeev:2007tn,Kharzeev:2007jp,Fukushima:2008xe,Kharzeev:2009fn}.
\vskip0.3cm
It is important to establish whether the CME explanation of charge
asymmetry fluctuations is the correct one.  First, it would be a
direct observation of a topological effect in QCD.  Second, the
magnitude of this effect in the chirally broken phase is expected to
be much smaller and hence the observation of the CME would manifest the restoration of 
chiral symmetry in the medium.  The effort of
quantifying the charge asymmetry fluctuations in QCD matter and of
examining alternative explanations and backgrounds has already begun\cite{Bzdak:2009fc,Nam:2009jb,Fukushima:2010vw,Gorsky:2010dr,KerenZur:2010zw,Fu:2010rs,Schlichting:2010na,Asakawa:2010bu,Voloshin:2010ut,Orlovsky:2010ga,Zhitnitsky:2010zx,Muller:2010jd,Mages:2010bc,Rogachevsky:2010ys}, 
and there are plans to further study this effect at RHIC, LHC, NICA and FAIR.
\vskip0.3cm
Recently, a new test of the chiral magnetic and chiral vortical effects (CME and CVE) has been proposed \cite{Kharzeev:2010gr}. The test relies only on the general properties of triangle anomalies. 
Consider anomalous
hydrodynamics~\cite{Son:2009tf}, and  suppose that the system under
consideration has a chemical potential $\mu$, coupled to a charge
$\bar q\gamma^0 B q$, where $B$ is a flavor matrix, and an axial
chemical potential $\mu_5$, coupled to the axial charge $\bar q
\gamma^0\gamma^5 A q$, where $A$ is another flavor matrix.  For
simplicity, we shall assume that both $\mu$ and $\mu_5$ are much
smaller than the temperature $T$ (this assumption usually holds in relativistic heavy ion collisions).  We also assume that electromagnetism
couples to the current $\bar q \gamma^\mu Q q$, with $Q$ being the
charge matrix.  If one measures a vector current $J^\mu = \bar
q\gamma^\mu V q$, then the result is
\begin{equation}
  \vec J = \frac{N_c\mu_5}{2\pi^2}\ [\tr (V\!AQ)\ \vec B + \tr(V\!AB)\ 2\mu \vec \omega]
\end{equation}
where $\vec B$ and $\vec \omega$ are the external magnetic fields and
the fluid vorticity respectively.  The two parts of the current on the right hand
side correspond to the CME and the CVE, respectively.  The traces in
the formula are related to the anomalous triangle diagram.
\vskip0.3cm
We shall consider two cases: $N_f=3$, where $u$, $d$ and $s$ quarks
are light, and $N_f=2$ where only $u$ and $d$ quarks are light.  In
both cases, we assume $A$ to be the unity matrix, $A=\dblone$ (which
is expected if the chiral asymmetry is due to instanton events, which
are flavor symmetric), and $B=(1/3)\dblone$.  For $N_f=3$, $Q={\rm
diag}(2/3,-1/3,-1/3)$, and for $N_f=2$, $Q={\rm diag}(2/3,-1/3)$.
There are two currents that we will measure: the electromagnetic
current $J_E$, corresponding to $V=Q$ and the baryon current $J_B$,
corresponding to $V=B$.
For CME, we get for the charge current (up to an overall factor of
$ N_c\ \mu_5 \vec B/(2\pi^2)$ which is common for both charge and baryon currents)
\beq
J_E^{CME} \sim \frac{2}{3} \ \  (N_f=3)\ \   {\rm or} \ \ \ \frac{5}{9} \ \ (N_f=2)
\eeq
and for the baryon current
\beq
J_B^{CME} = 0 \ \  (N_f=3)\ \   {\rm or} \ \ \ \sim \frac{1}{9} \ \ (N_f=2).
\eeq
For CVE, the results are (up to the overall factor $N_c\ \mu_5\mu \vec\omega/\pi^2$)
\beq
J_E^{CVE} = 0  \ \  (N_f=3)\ \   {\rm or} \ \ \ \sim \frac{1}{3} \ \ (N_f=2);
\eeq
\beq
J_B^{CVE} \sim 1 \ \  (N_f=3)\ \   {\rm or} \ \ \ \sim \frac{2}{3} \ \ (N_f=2).
\eeq

In the SU(3) case, the CME and CVE lead to completely different
currents: the CME contributes only to the electromagnetic current and the
CVE contributes only to the baryon current.  In the SU(2) case, the
separation is less clean, but the ratio of $J_B/J_E$ still differs by a
factor of ten.
\vskip0.3cm
Let us now discuss the implications for heavy ion
collisions.  It is known that the baryon chemical potential of the produced fireball
depends on the collision energy: at smaller $\sqrt s$, $\mu$ is
larger.  Thus the CVE should be more important at lower energies.  According
to the computation above, $J_B/J_E$ becomes larger as one lowers the
energy of the collision.  Moreover, since the symmetry arguments suggest that the magnetic
field and the vorticity of the fluid have to be aligned, our results show that the two vectors
$\vec J_B$ and $\vec J_E$ should point in the same direction.
Therefore, in addition to the charge separation,
there must be a baryon number separation.  The two effects are
positively correlated on the event-by-event basis, and the relative
importance of baryon number separation increases as one lowers the
energy of the collision.

\section{Summary}

The picture proposed by Gribov identifies the high-momentum collective motion in the Dirac sea as a source 
of axial anomaly. In a strong magnetic field, and in the presence of a local chirality imbalance, this collective motion becomes directly observable in the form of electric CME current. In heavy ion collisions, the local chirality imbalance can be readily supplied by the topological gluon field configurations in hot QCD matter, and a sufficiently strong magnetic field is delivered by the colliding ions. The effect thus can become observable, and there is an intriguing evidence from RHIC experiments for the expected charge asymmetry fluctuations\cite{:2009txa,phenix}.  Much remains to be done to verify (or disprove) the anomaly-related origin of the observed effect; however this work has already begun. 
\vskip0.3cm
I am indebted to Julia Nyiri and Yuri Dokshitzer for their kind invitation to the Gribov-80 Conference. I thank my collaborators G. Ba\c sar, \mbox{P. Buividovich,} M. Chernodub, G. Dunne, K. Fukushima, L. McLerran, \mbox{M. Polikarpov,} D. Son, \mbox{H. Warringa,} H.-U. Yee, and A. Zhitnitsky for sharing their insights with me and numerous enjoyable discussions. 
This work was supported in part by the U.S. Department of Energy 
under Contract No.~DE-AC02-98CH10886.
\bibliographystyle{ws-procs9x6}
\bibliography{ws-pro-sample}

\end{document}